\documentclass[aps,prd,reprint,preprintnumbers,floats,epsfig,nofootinbib,amssymb,
nofootinbib]{revtex4}

\usepackage{slashed}
\usepackage{graphicx,color}
\usepackage{epsfig}
\usepackage{subfigure}
\usepackage{epsfig}
\usepackage{epstopdf}
\usepackage{dcolumn}
\usepackage{multirow}
\usepackage{bm}
\usepackage{color}
\usepackage{ulem}
\usepackage{enumitem}
\usepackage{amsmath}
\usepackage[colorlinks,
linkcolor=black,
anchorcolor=black,
citecolor=black
]{hyperref}

\begin{document}

	\baselineskip=15pt
	
	\preprint{CTPU-PTC-25-26}

\title{
Probing maximal flavor changing $Z'$ in $U(1)_{L_\mu-L_\tau}$   at $\mu$TRISTAN
}

\author{Fei Huang${}^{1}$\footnote{sps\_huangf@ujn.edu.cn}}
\author{ Jin Sun$^{2}$\footnote{sunjin0810@ibs.re.kr(contact author)}}

\affiliation{${}^{1}$School of Physics and Technology, University of Jinan, Jinan, Shandong 250022, China}

\affiliation{${}^{2}$Particle Theory and Cosmology Group, Center for Theoretical Physics of the Universe, Institute for Basic Science (IBS), Daejeon 34126, Korea}

\begin{abstract}

We explore  the potential to detect  the $U(1)_{L_\mu-L_\tau}$ model featuring triplet scalars $\Delta$ at the $\mu$TRISTAN collider. 
The new gauge boson $Z'$, arising from the spontaneous breaking of $U(1)_{L_\mu-L_\tau}$, can exhibit maximal flavor changing interactions under the exchange symmetry, while $\Delta$ mediates the flavor conserving interactions.
The absence of muon $(g-2)_\mu$  
can be explained by interference effects arising from opposite contributions of $Z'$ and $\Delta$, with similar interference patterns also manifesting in the tau decay process $\tau\to \mu\nu\bar\nu$.
These counteracting effects render the model phenomenologically interesting and warrant further investigation.
For the mass $m_{Z'}$  in the range of hundreds of GeV, we find that   $\mu^+\mu^+$ and $\mu^+e^-$ collider at the $\mu$TRISTAN
 can probe many regions inaccessible to current experiments and offer greater projected sensitivity than opposite-sign muon colliders.
This suggests that $\mu$TRISTAN can serve as complementary exploration to the $U(1)_{L_\mu-L_\tau}$ model,  providing compelling motivation for  the next generation of high-energy lepton colliders.

\end{abstract}

\maketitle

\section{Introduction}

High-energy muon colliders have consistently attracted significant attention over the past decades for studying the Standard Model (SM) and searching for new physics, owing to their high center-of-mass energy and clean lepton collider environment. 
To date, most muon collider studies  have focused on the opposite-sign muon cases~\cite{Han:2022edd,Azatov:2022itm,Bredt:2022dmm,Arakawa:2022mkr,Lv:2022pts,Reuter:2022zuv,Kwok:2023dck,Li:2023tbx,Li:2023ksw,Inan:2023pva,Lu:2023ryd,Maharathy:2023dtp,Reuter:2023ivd,Amarkhail:2023xsc,Belyaev:2023yym,Yang:2023gos,Belfkir:2023lot,Forslund:2023reu,Loisa:2023djb,	Liu:2023yrb,Sun:2023ylp,Ouazghour:2023plc,Ghosh:2023xbj,Mikulenko:2023ezx}.
The possibility of exploring physics with same-sign muon colliders~\cite{Heusch:1995yw}, referred to as $\mu$TRISTAN~\cite{Hamada:2022mua}, has recently gained attention.
This collider utilizes a low-emittance muon beam originally developed for the muon $g-2$/EDM experiment at J-PARC~\cite{Abe:2019thb}. 
It can accelerate $\mu^+$ and $e^-$ beams up to 1 TeV and 30 GeV, respectively, resulting in  the center-of-mass
energy of  346 GeV for $\mu^+e^-$ and 2 TeV for $\mu^+\mu^+$ collider experiments.
The luminosities are estimated to be at the
level of 10$^{33-34}$ cm$^{-2}$s$^{-1}$~\cite{Hamada:2022mua,Hamada:2022uyn}, which makes $\mu^+e^-$ collider as an excellent Higgs boson factory and high energy  $\mu^+\mu^+$  collider as a promising facility for direct searches of new particles~\cite{Jiang:2023mte,Li:2023lkl,Yang:2023ojm,Fukuda:2023yui,Dev:2023nha,Goudelis:2023yni,Das:2024gfg,Lichtenstein:2023iut}.

The detection capabilities of opposite-sign and same-sign muon colliders can be compared in a model-independent manner. 	The effective four-fermion operators are
		\begin{eqnarray}
			{\cal L} =	C_{AB}^{ijkl}(\bar l_i \gamma^\mu P_A l_j)(\bar l_k \gamma^\mu P_B l_l)+h.c. \;.
		\end{eqnarray}
		Here A and B means the chirality projection  $P_{L/R}=(1\mp \gamma_5)/2$, and $i, j, k, l$ are indices of charged lepton flavors $e,\mu, \tau$.
         Take $\mu \mu \to \tau \tau$ as an example to illustrate, the coefficient should be $C_{AB}^{\tau\mu\tau\mu}$.
         This leads to the ratio of the cross section at $\mu^+\mu^+$ and $\mu^+\mu^-$ colliders,  expressed as~\cite{Fridell:2023gjx} 
		\begin{eqnarray}\label{EFT}
			\frac{\sigma(\mu^+\mu^+\to \tau^+\tau^+)}{\sigma(\mu^+\mu^-\to \tau^+\tau^-)}
			=3\times \frac{|C_{LL}|^2+ |C_{RR}|^2+(|C_{LR}|^2+ |C_{RL}|^2)/6}{|C_{LL}|^2+ |C_{RR}|^2+|C_{LR}|^2+ |C_{RL}|^2} \;.
		\end{eqnarray}
For simplicity, we omit the superscripts of flavor indices. This clearly indicates that in models where $C_{LL}$ or $C_{RR} \neq 0$, the sensitivity of the $\mu^+ \mu^+$ collider could surpass that of the $\mu^+ \mu^-$ collider, emphasizing the necessity of analyses involving the $\mu^+ \mu^+$ collider. 

Recently, FNAL released the latest precise measurements of the muon magnetic moment and combined them with previous results~\cite{Muong-2:2021ojo,Muong-2:2023cdq} to obtain a new experimental world average, $a_\mu^{exp}=1165920715(145)\times 10^{-12}$~\cite{Muong-2:2025xyk}.
Additionally, the theory community has updated the Standard Model (SM) prediction by considering only lattice-QCD calculations, excluding the data-driven dispersive approach, yielding $a_\mu^{SM}=116 592 033(62)\times 10^{-11}$~\cite{Aliberti:2025beg}.	
The deviation is $a_\mu^{exp}-a_\mu^{SM}=39(64)\times 10^{-11}$, indicating no significant tension between the SM and experimental results at the current precision level.
 However, the result requires further analysis due to large uncertainties, and theoretical discrepancies between lattice-QCD calculations and the data-driven dispersive approach.
Nevertheless, it also imposes stringent constraints on any new model building.
A widely known model, the $L_\mu-L_\tau$ model, is subject to stringent bounds from various experiments in the low-mass region ($m_{Z'} < 70$ GeV), including BaBar for $m_{Z'} \sim (0.2,10)$ GeV~\cite{Godang:2016gna,Capdevilla:2021kcf}, CMS $Z \to 4\mu$ for $m_{Z'} \sim (5,70)$ GeV~\cite{CMS:2012bw,Altmannshofer:2014pba,CMS:2018yxg,Capdevilla:2021kcf}, BOREXINO for $m_{Z'} \sim (0.01,1)$ GeV~\cite{Kamada:2015era,Gninenko:2020xys}, and NA64$\mu$~\cite{NA64:2024klw} below 1 GeV.
Given the stringent collider bounds in the low-mass region, we focus on the high-mass regime (on the order of hundreds of GeV), where the primary relevant constraints arise from muon neutrino trident (MNT) production~\cite{Altmannshofer:2014pba,Cen:2021ryk}.
Fortunately, previous studies~\cite{He:1990pn,Foot:1994vd,Cheng:2021okr,Huang:2024iip} indicate that the MNT bound can be evaded by transforming the original flavor-conserving $Z'$ interaction into a fully off-diagonal form with the assistance of type-II triplets and exchange symmetry.
Interestingly, the physical effects of the maximally flavor-violating $Z'$ and the triplet scalars are mutually opposite. For example, the $Z'$ provides a positive contribution to $a_\mu$, whereas the triplet contributes negatively, effectively reinforcing the latest results about the absence.
These mutually counteracting effects make the new model mechanism worthy of further study.
Previous collider analyses have focused on the opposite-sign muon $\mu^+ \mu^-$ collider, including the two-body process $\mu^- \mu^+ \to \tau^- \tau^+$ and the four-body process  $\mu^-\mu^+ \to \mu^\pm\mu^\pm + \tau^\mp \tau^\mp$~\cite{Sun:2023ylp}.
The new model mechanism can be advantageously explored at the $\mu$TRISTAN collider, as analyzed in Eq.~(\ref{EFT}).

In this paper, we aim to utilize the $\mu^+\mu^+$ and $\mu^+e^-$ colliders to search for a maximally flavor-changing $Z'$ boson within the $L_\mu-L_\tau$ model.
The structure of the paper is organized as follows. 
In section \ref{sec2}, we discuss the model construction of  the maximal flavor changing $L_\mu-L_\tau$ $Z'$ interactions 
 augmented by triplet scalars.
Section \ref{sec3} presents the collider signatures,  firstly analyzing the two-body case in $\mu^+\mu^+$ collider to fix the triplet effects, followed by an extension to the four-body processes at both the $\mu^+\mu^+$ and $\mu^+e^-$ colliders. 
We provide our conclusions in section \ref{sec4}.


\section{The $U(1)_{L_{\mu}-L_{\tau}}$ model for maximal $\mu$-$\tau$ coupling}
\label{sec2}

In the minimal $U(1)_{L_\mu - L_\tau}$ model, the left-handed $SU(3)_C\times SU(2)_L\times U(1)_Y$ doublets, $L_{L\;i}: (1, 2, -1/2)$, and the right-handed singlets, $e_{R\;i}: (1,1,-1)$, transform under the gauged $U(1)_{L_\mu-L_\tau}$ group with charges $0$, $+1$, and $-1$ for the first, second, and third generations, respectively. The model’s $Z^\prime$ gauge boson interacts exclusively with leptons in the weak interaction basis~\cite{He:1990pn, He:1991qd}
\begin{eqnarray}\label{conserving}
	{\cal L}_{Z'}=- \tilde g (\bar \mu \gamma^\mu \mu - \bar \tau  \gamma^\mu \tau + \bar \nu_\mu \gamma^\mu P_L \nu_\mu - \bar \nu_\tau \gamma^\mu P_L \nu_\tau) Z^\prime_\mu \;, \label{zprime-current}
\end{eqnarray}
where $\tilde g$ is the $U(1)_{L_\mu - L_\tau}$ gauge coupling.
By introducing a singlet scalar field $S$ with vacuum expectation value (VEV) $v_S/\sqrt{2}$ and a $U(1)_{L_\mu - L_\tau}$ charge  $+1$ to spontaneously break the $U(1)_{L_\mu - L_\tau}$ symmetry, the $Z'$ boson acquires a mass given by $m_{Z'} = \tilde g v_S$.

The above interaction is purely flavor-conserving but can be transformed into a fully flavor-changing interaction. We briefly outline the construction of such a model. This requires introducing three Higgs doublets, $H_{1,2,3}: (1,2, 1/2)$ with vevs $<H_i> = v_{i}/\sqrt{2}$, carrying $U(1)_{L_\mu - L_\tau}$ charges of $0$, $+2$, and $-2$, respectively. An unbroken exchange symmetry is imposed such that $Z^\prime \to - Z^\prime$,  $H_1 \leftrightarrow H_1$ and $H_2 \leftrightarrow H_3$ with $v_2 = v_3 = v$. Under these conditions, the $Z'$ interactions and Yukawa couplings to leptons are given by
\begin{eqnarray}\label{yukawa}
	L _{H}= &&- \tilde g (\bar l_2 \gamma^\mu L l_2- \bar l_3  \gamma^\mu L l_3 + \bar e_2 \gamma^\mu R e_2 - \bar e_3 \gamma^\mu R e_3) Z^\prime_\mu\nonumber\\
	&&- [Y^l_{11} \bar l_1 R e_1 + Y^l_{22} (\bar l_2 R e_2 +\bar l_3 R e_3 ) ] H_1 
	-Y^l_{23} (\bar l_2 R e_3 H_2 +\bar l_3 R e_2 H_3 ) + H.C.
\end{eqnarray}
After electroweak symmetry breaking, the scalars acquire non-zero  VEVs that generate the charged lepton masses through the Yukawa terms shown in the second line of Eq.~(\ref{yukawa}). However, since the resulting mass matrix is non-diagonal, it must be diagonalized via the following transformation:
\begin{eqnarray}\label{eigen}
	\left (
	\begin{array}{c}
		\mu\\
		\tau
	\end{array}
	\right )
	= \frac{1}{\sqrt{2}}
	\left (
	\begin{array}{rr}
		1&\;-1\\
		1&\;1
	\end{array}
	\right )
	\left (\begin{array}{c}
		e_2\\
		e_3
	\end{array}
	\right ),\;
	\end{eqnarray}
After applying the above transformation to the charged lepton mass eigenstate basis, we obtain the desired flavor-changing $Z^\prime$ interactions in the form  as 
	\begin{eqnarray}
		{\cal L}_{Z'}=	- \tilde g (\bar \mu \gamma^\mu \tau +  \bar \tau  \gamma^\mu \mu + \bar \nu_\mu \gamma^\mu L \nu_\tau  + \bar \nu_\tau \gamma^\mu L \nu_\mu) Z'_\mu \;. \label{zprime-changing}
	\end{eqnarray}
The aforementioned $Z'$ interaction also induces the process $\tau \to \mu \bar \nu_\mu \nu_\tau$, whose effects can be mitigated by introducing three $Y=1$ scalar triplets, $\Delta_{1,2,3}: (1,3,1)$ ($<\Delta_i> =  v_{\Delta i}/\sqrt{2}$)  with $U(1)_{L_\mu-L_\tau}$ charges ($0,-2,2$)~\cite{Cheng:2021okr}.   
This $\Delta$ field is core of the famous  type-II seesaw mechanism providing small neutrino masses~\cite{Lazarides:1980nt,Mohapatra:1980yp,Konetschny:1977bn,Cheng:1980qt,Magg:1980ut,Schechter:1980gr} with 
	\begin{eqnarray}
		\Delta = \left (\begin{array}{cc}   \Delta^+/\sqrt{2}&\;  \Delta^{++}\\  \Delta^0&\; - \Delta^+/\sqrt{2} \end{array} \right )\;,\;\;\;\Delta^0=\frac{v_\Delta+\delta+i\eta}{\sqrt{2}}\;. 
	\end{eqnarray}
Under the above exchange symmetry $\Delta_1 \leftrightarrow  \Delta_1,\; \Delta_2 \leftrightarrow  \Delta_3$ with $v_{\Delta 2} = v_{\Delta 3}$, we use the transformation  in Eq.(\ref{eigen}) to obtain the Yukawa terms in terms of the component fields $\Delta^{0, +, ++}$ as
	\begin{eqnarray}
		&&
		L_\Delta = - (\bar \nu_e^c, \bar \nu_\mu^c, \bar \nu_\tau^c ) M(\Delta^0)  L
		\left ( \begin{array}{c}
			\nu_e \\ \nu_\mu \\ \nu_\tau
		\end{array}
		\right )
		+ \sqrt{2} (\bar \nu_e^c, \bar \nu_\mu^c, \bar \nu_\tau^c ) M(\Delta^+) L
		\left ( \begin{array}{c}
			e \\ \mu \\ \tau
		\end{array} \right )
		+ (\bar e^c, \bar \mu^c, \bar \tau^c ) M(\Delta^{++}) L
		\left ( \begin{array}{c}
			e \\ \mu \\ \tau
		\end{array} \right )
		\;,\nonumber\\
		&&\mbox{with}\;\;\; M(\Delta)  = \left ( \begin{array}{ccc}
			Y^\nu_{11} \Delta_1&0&0\\
			\\
			0& (Y^\nu_{22}(\Delta_2 +\Delta_3) - 2 Y^\nu_{23}\Delta_1)/2 & Y^\nu_{22}(\Delta_2-\Delta_3)/2\\
			\\
			0&Y^\nu_{22}(\Delta_2-\Delta_3)/2&(Y^\nu_{22}(\Delta_2 +\Delta_3) + 2 Y^\nu_{23}\Delta_1)/2
		\end{array}
		\right ) \;.
	\end{eqnarray}
 We find that numerous model parameters complicate the analysis. For simplicity, we make the following assumptions:
\begin{itemize}
\item \textbf{ $\Delta_2=\Delta_3$}: 
The degenerate case is enforced by the exchange symmetry, which forbids off-diagonal interactions.
\item \textbf{$Y_{11,23}<<Y_{22}$}: 
This suppresses the effects of $\Delta_1$, leaving the triplet contributions dominated by $\Delta_{2}$ and $\Delta_{3}$.
\item \textbf{$ m_{\Delta^{++,+,0}}= m_\Delta$}: The degeneracy of the triplet components ensures that the triplet scalars have identical masses.
\end{itemize}
The above assumptions ensure that the triplet effects can be effectively described by the coupling $y = Y_{22}$ and the mass parameter $m_\Delta = m_{\Delta_2} = m_{\Delta_3}$.

\section{collider signature analysis}
		\label{sec3}

Note that  our models include both  the 
fully flavor-changing $Z'$ interactions and flavor-conserving $\Delta$ interactions, characterized by four parameters: $\tilde{g}$, $m_{Z'}$, $y$, and $m_\Delta$.
However, the combined effects of these four parameters contribute to various physical processes, making it difficult to isolate the individual contributions from $Z'$ or $\Delta$.
In this section, we first analyze the decay chains of $\Delta$ and $Z'$ to identify their dominant decay channels. Then, we investigate the physical effects of the triplet scalars through the specific process $\mu^+\mu^+\to W^+W^+, \mu^+\mu^+$. Finally, we extend the analysis to four-body collider processes involving both $Z'$ and $\Delta$.

\subsection{decay chains of triplet $\Delta^{++}$ and $Z'$ bosons}

\begin{figure*}
    \centering
    \subfigure[\label{fig:X}]
    {\includegraphics[width=0.486\linewidth]{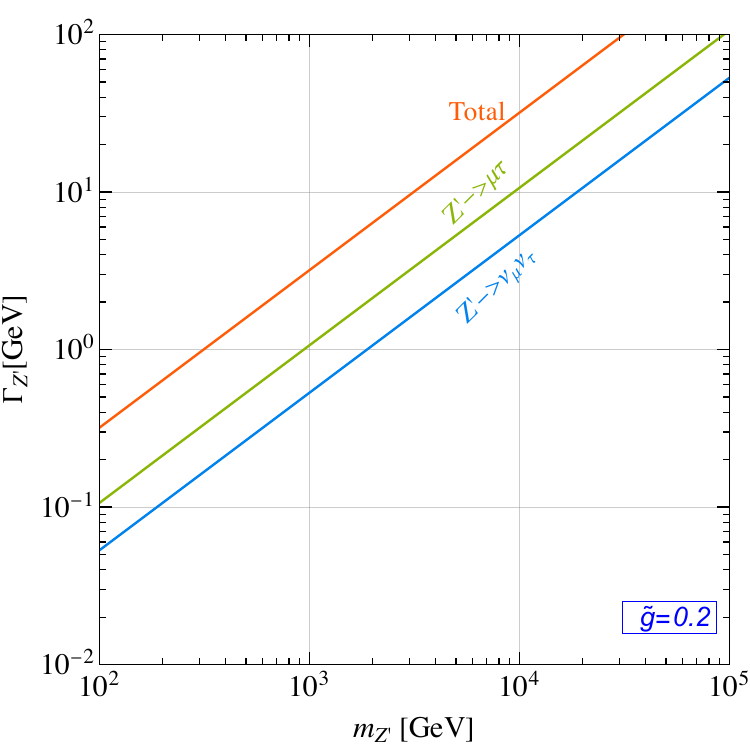}}
    \subfigure[\label{fig:W}]
    {\includegraphics[width=0.486\linewidth]{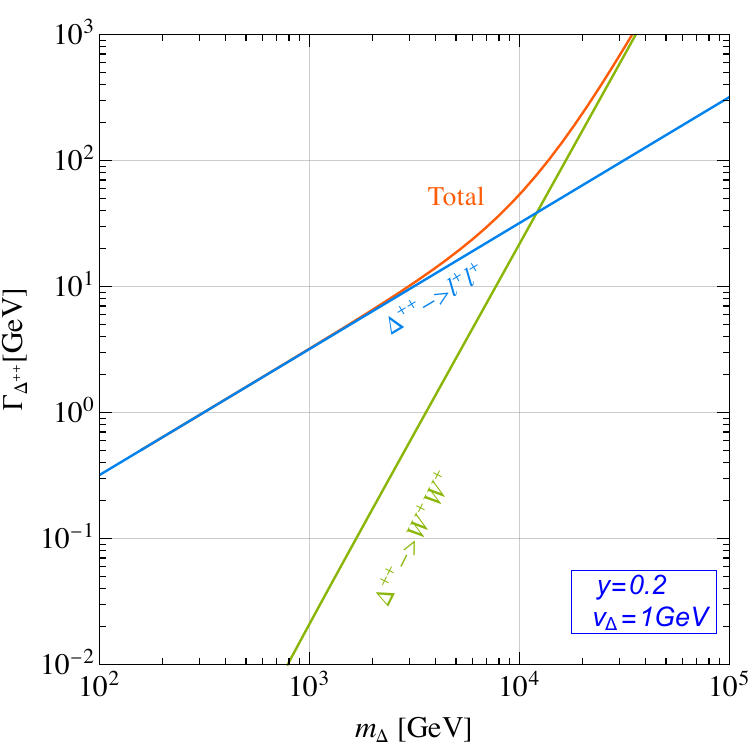}}
    \caption{The decay width as functions of mass  in our model. Different decay channels are distinguished by distinct colors.
    (a) The left panel corresponds to $Z'$ decay width for $\tilde g=0.2$.
    (b) The right panel means  triplet scalar $\Delta^{++}$ decay width for $y=0.2$ and $v_\Delta=1$ GeV. 
   }
    \label{fig:width}
\end{figure*}

For the $Z'$ gauge boson, it only mediates the flavor changing $\mu-\tau$ interactions, 
leading to decay chains involving charged leptons and neutrinos. The relevant decay widths are calculated as
\begin{eqnarray}
&&\Gamma_{Z'}=\Gamma(Z'\to\mu^+\tau^-)+\Gamma(Z'\to\mu^-\tau^+)+\Gamma(Z'\to \nu_{\mu}\bar \nu_{\tau})+ \Gamma(Z'\to \nu_{\tau}\bar \nu_{\mu})\;,\nonumber\\
&&\Gamma(Z'\to\mu^\pm\tau^\mp)=\frac{\tilde g^2}{24\pi}\frac{1}{m_{Z'}^5}(m_{Z'}^2-m_{\tau}^2)^2(2m_{Z'}^2+m_{\tau}^2)\;,\quad
 \Gamma(Z'\to \nu_{\mu(\tau)}\bar \nu_{\tau(\mu)})=\frac{\tilde g^2m_{Z'}}{24\pi}\;.
\end{eqnarray}
Here, we neglect the muon mass. In the limit $m_{Z'} \gg m_\tau$, the decay width into neutrinos is half that of the charged lepton decay, due to the involvement of only left-handed neutrinos. Consequently, the branching ratio satisfies $Br(Z' \to \text{invisible}) \approx 2/3$, provided the decay channel $Z' \to \mu \tau$ is kinematically allowed. 
The corresponding decay widths are presented in Fig.~\ref{fig:X}, with the coupling fixed at $\tilde{g} = 0.2$. For different coupling constants, the decay width can be rescaled as  $\Gamma=(\tilde g/0.2)^2\Gamma(\tilde g=0.2)$. This indicates that increasing the coupling $\tilde{g}$ and the mass $m_{Z'}$ both enhance the decay width.

For triplet scalars $\Delta$, 
the degeneracy of triplet masses forbids decay processes between different triplet components. This implies that the effects can be probed via the decay of the doubly-charged scalar $\Delta^{++}$ into SM particles through two distinct decay channels, as follows
    \begin{eqnarray}
   &&  \Gamma_\Delta =\Gamma(\Delta^{++}\to W^+W^+)+\Gamma(\Delta^{++}\to \ell^+\ell^+)\;,\\
    &&   \Gamma(\Delta^{++}\to W^+W^+) = \frac{G_F^2  v_\Delta^2}{2\pi m_\Delta} [(m_\Delta^2-2m_W^2)^2+8m_W^4] 
    	\sqrt{1-\frac{4m_W^2}{m^2_\Delta}}\;,\quad  \Gamma(\Delta^{++}\to \ell^+\ell^+) =
        \sum_{\mu,\tau}\frac{y^2m_\Delta}{8\pi}\sqrt{1-\frac{4m_l^2}{m^2_\Delta}}\;,\nonumber
    \end{eqnarray}
where $G_F=1.1664\times 10^{-5}\mbox{GeV}^{-2}$  is Fermi constant.
The decay channel into leptons is proportional to the Yukawa coupling, whereas the decay channel into two $W$ bosons depends on vev. Comparing these two decay modes, we find that the ratio of their decay widths is given by
 \begin{eqnarray}
     \frac{\Gamma(\Delta^{++}\to W^+W^+)}{\Gamma(\Delta^{++}\to \ell^+\ell^+)} \approx \frac{v_\Delta^2}{v_{SM}^4}\frac{ m_\Delta^2 }{ y^2}\approx 10^{-10}\left(\frac{v_\Delta}{1\text{GeV}}\right)^2\left(\frac{250\text{GeV}}{v_{SM}}\right)^4\left(\frac{m_\Delta/\text{GeV}}{y}\right)^2\;.
 \end{eqnarray}
 here we use $G_F=1/(\sqrt{2}v^2_{SM})$ with $v_{SM}=246$GeV.  
The corresponding decay widths are presented in Fig.~\ref{fig:W}, where we set $y=0.2$ and $v_\Delta = 1$ GeV. For $m_\Delta$ below 10 TeV, the dominant decay channel is into charged leptons, $\Delta^{++}\to \ell^+\ell^+$. Conversely, for $m_\Delta > 10$ TeV, the dominant decay channel shifts to $\Delta^{++} \to W^+ W^+$. 
Increasing $v_\Delta$ or $m_\Delta$, or decreasing $y$, enhances the decay into the $W W$ channel. 
Combining the decay widths of both $Z'$ and $\Delta^{++}$, we find that the widths are generally of order 1 GeV for masses in the hundreds of GeV range. This corresponds to a decay length of approximately 1 fm, indicating that both $Z'$ and $\Delta^{++}$ undergo prompt decays.

Before performing a detailed collider analysis, we first examine the corresponding constraints on the model parameters.
\begin{itemize}
\item $(g-2)_l$: Exchanging $\Delta^{+,++}$ at one loop level can contribute to electron $(g-2)_l$, while 
$Z'$ and $\Delta^{++,+}$
provide opposite contributions, expressed as~\cite{Cheng:2021okr,Huang:2024iip}
\begin{eqnarray}\label{largeg-2}
\Delta a_{e}^{\Delta}
 =-\frac{Y_{11}^2}{16\pi^2}\frac{m_{e}^2}{m_{\Delta_1}^2}\;,\quad 
\Delta a_{\mu}=\Delta a^{Z'}_{\mu}+ 
\Delta a^{\Delta}_{\mu}
=\frac{\tilde g_{Z'}^2 m_{\mu}^2}{12\pi^2m_{Z'}^2}\left(\frac{3m_{\tau}}{m_{\mu}}-2\right) -\frac{m_{\mu}^2}{16\pi^2}
   \frac{y_{22}^2}{m_{\Delta}^2}\;.
\end{eqnarray}
Here we adopt the limit $m_{\Delta,Z'}>>m_{\tau,\mu}$.
For $(g-2)_e$, 
the triplet scalar provides a negative contribution consistent with the measured value $\Delta a_e(Cs)=(-101+27)\times 10^{-14}$~\cite{Parker:2018vye}. Although our analysis neglects the parameters  $Y_{11}$ and $\Delta_1$ for simplicity, viable parameter regions can still accommodate $(g-2)_e$. Furthermore, the opposite contributions from $Z'$ and $\Delta$ mutually cancel to comply with the recent muon $(g-2)_\mu$ measurement, $a_\mu = 39(64) \times 10^{-11}$. We apply a $2\sigma$ confidence level to impose exclusion limits on the model parameters in the $\epsilon - m_{Z'}$ plane, shown in the upper left corner of Fig.~\ref{fig:all}. 
\item \textbf{$\tau \to \mu\nu\bar\nu$}: The decay processes are mediated by $Z'$ and $\Delta^+$ effects in terms of the ratios as~\cite{Cheng:2021okr,Huang:2024iip}
\begin{eqnarray}
     R^\tau&\equiv&\frac{\Gamma(\tau^-\to\mu^-\overline{\nu}_\mu\nu_\tau)}{\Gamma(\tau^-\to\mu^-\overline{\nu}_\mu\nu_\tau)_{\mathrm{SM}}}=\left(1-\frac{4m_W^2}{g^2}\frac{y_{22}^2}{4m_{\Delta^+}^2}\right)^2+\frac{4m_W^2}{g^2}\frac{\tilde g^2}{m_{Z^{\prime}}^2}\left(1-\frac{4m_W^2}{g^2}\frac{y_{22}^2}{4m_{\Delta^+}^2}
  \right)+ \frac{\tilde g^4}{g^4}\frac{4m_W^4}{m_{Z^{\prime}}^4}\;.
\end{eqnarray}
Combining the  SM predictions $R^\tau_{SM}= 0.972564\pm0.00001$~\cite{Pich:2009zza} and average experimental values $R^\tau_{aver}=0.972968\pm 0.002233$ from HFLAV collaboration~\cite{HFLAV:2022esi} and Belle-II~\cite{Corona:2024nnn}, the ratio can be obtained as  $R^\tau=1.00042\pm 0.00230$, which imposes the stringent bounds as shown in the  Fig.~\ref{fig:all}.
\item \textbf{collider bounds}: 
Current constraints on the degenerate triplet scalar set a lower mass limit of $m_\Delta > 420$ GeV for $\Delta m = 0$ and $v_\Delta \sim \mathcal{O}(\mathrm{GeV})$~\cite{Ashanujjaman:2021txz}.
\item \textbf{Neutrino trident processes}:
The process  $\nu_\mu+N\to \nu_{\mu,\tau} +N +\mu^+\mu^-$ can be induced by $\Delta^+$ at tree-level as~\cite{Cheng:2021okr,Huang:2024iip} 
\begin{eqnarray}\label{sigmaDelta^+}
   \frac{\sigma_{\Delta^+}}{  \sigma_{SM}}|_{trident}=
   \frac{\left(1+4 s^2_W- \frac{m^2_W}{g^2}\frac{2y_{22}^2}{m_{\Delta}^2}\right)^2+ 
    \left(1- \frac{m^2_W}{g^2}\frac{2y_{22}^2}{m_{\Delta}^2}\right)^2} 
   {[(1+4 s^2_W)^2 + 1]}\;,
\end{eqnarray}
Combining the current bounds from CHARM, CCFR and NuTeV~\cite{CHARM-II:1990dvf,CCFR:1991lpl,NuTeV:1999wlw}, the averaged value is $0.95\pm0.25$, which can place an upper bound 
 $y^2/m_{\Delta}^2<2\times 10^{-5}$ GeV$^{-2}$.
\end{itemize}

\subsection{two-body final states }

Given the interplay between the two new particles ($Z'$ and $\Delta$), we aim to disentangle their overlapping effects by analyzing distinct scattering processes. 
In certain processes, only the triplet scalar $\Delta$ contributes, while $Z'$ does not, such as  $\mu^+\mu^+\to W^+W^+$ and $\mu^+\mu^+\to \mu^+\mu^+$.  Therefore, these two processes provide an ideal platform to constrain the viable parameter space of the triplet $\Delta$.

\subsubsection{Identify the triplet effects}

In order isolate the effects of the triplet scalar, we focus on processes solely mediated by it, specifically $\mu^+\mu^+ \to W^+ W^+$ and $\mu^+\mu^+ \to \mu^+\mu^+$. 

\begin{figure}
    \centering
\includegraphics[width=0.8\linewidth]{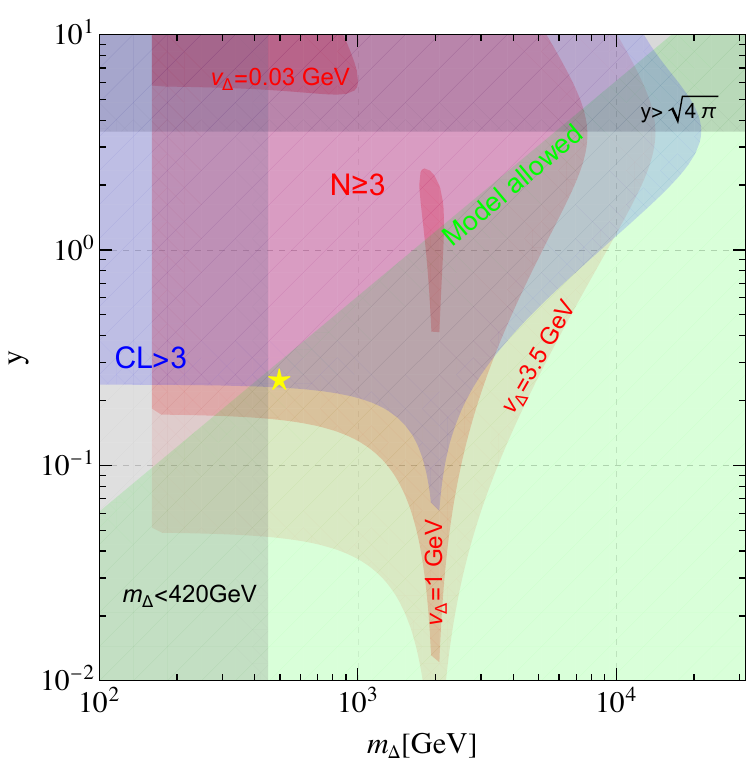}
    \caption{
    The allowed parameter space for the triplet scalars  in the $m_\Delta$–$y$ plane.
Different constraints are indicated using distinct colors: green for phenomenological limits (from $(g-2)_\mu$, MNT production, and $\tau \to \mu \nu \bar\nu$)~\cite{Huang:2024iip}; red for regions satisfying $N(\mu^+ \mu^+ \to W^+ W^+) \geq 3$ under three representative values of $v_\Delta = (0.03, 1, 3.5)$ GeV; blue for $CL(\mu^+ \mu^+ \to \mu^+ \mu^+) \geq 3$ as defined in Eq.(\ref{crit}); and gray for excluded regions due to the perturbativity condition $y \leq \sqrt{4\pi}$ and the lower bound on the degenerate triplet scalar mass $m_\Delta > 420$ GeV\cite{Ashanujjaman:2021txz}. The yellow star denotes the chosen benchmark point: $y = 0.25$, $m_\Delta = 500$ GeV, and $v_\Delta = 1$ GeV.
   }
    \label{fig:bound}
\end{figure}

In the $\mu^+\mu^+ \to W^+ W^+$ process, the mediation occurs exclusively via the $s$-channel exchange of the doubly-charged scalar $\Delta^{++}$. The corresponding cross section is computed as follows~\cite{Fridell:2023gjx}
     \begin{eqnarray}
    	\sigma(\mu^+ \mu^+ \to W^+ W^+) =\frac{G_F^2 y^2v_\Delta^2}{2\pi} \frac{(s-2m_W^2)^2+8m_W^4}{(s-m_\Delta^2)^2+m_\Delta^2 \Gamma_\Delta^2}
    	\sqrt{1-\frac{4m_W^2}{s}}\;,
    \end{eqnarray}
Here, $s$ denotes the center-of-mass (COM) energy. 
The cross section is found to depend on the $v_\Delta$, the mass $m_\Delta$, and the coupling constant $y$.
Assuming an integrated luminosity of $\mathcal{L}=12 fb^{-1}$~\cite{Hamada:2022mua}~\cite{Hamada:2022mua}, we derive constraints by requiring the number of signal events to satisfy $N_{sig}=\mathcal{L}\times\sigma \geq 3$.
The corresponding bounds are illustrated in Fig.\ref{fig:bound}. 
The green regions represent phenomenological constraints on the model parameters~\cite{Cheng:2021okr,Huang:2024iip}, including those from $(g-2)_{\mu}$,  MNT production, and $\tau \to \mu \nu \nu$ decay. 
The gray regions indicate limits arising from the degenerate triplet scalar mass bound $m\Delta > 420$ GeV~\cite{Ashanujjaman:2021txz} and perturbativity constraint $y \leq \sqrt{4\pi}$. We analyze three distinct scenarios corresponding to triplet VEV values of $v_\Delta = 0.03$, $0.5$, and $3.5$ GeV, marked by the red regions. Our results show that, to observe a viable signal, the triplet VEV should lie within the range $(0.03 - 3.5)$ GeV. The upper limit of $3.5$ GeV is set by the electroweak $\rho$ parameter at the 3$\sigma$ confidence level~\cite{PDG}. For $v_\Delta = 0.03$ GeV, the allowed triplet scalar mass peaks near $m_\Delta \sim \sqrt{s} = 2$ TeV. To streamline subsequent analyses of various collider processes, we select benchmark parameters indicated by a yellow star: $y = 0.25$, $m_\Delta = 500$ GeV, and $v_\Delta = 1$ GeV.

\begin{figure*}
    \centering
    \subfigure[\label{fig:mumu}]
     {\includegraphics[width=0.486\linewidth]{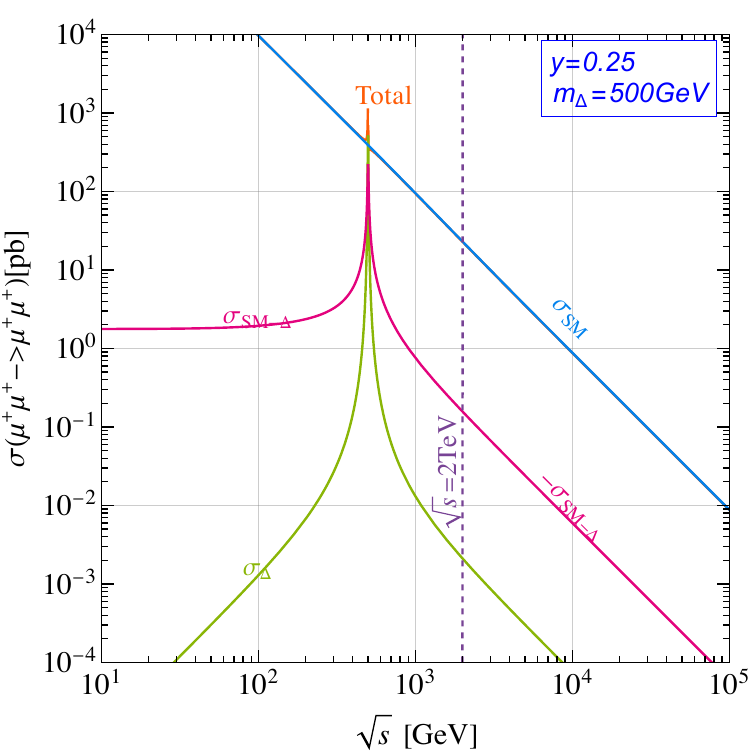}}
     \subfigure[\label{fig:tautau}]
    {\includegraphics[width=0.486\linewidth]{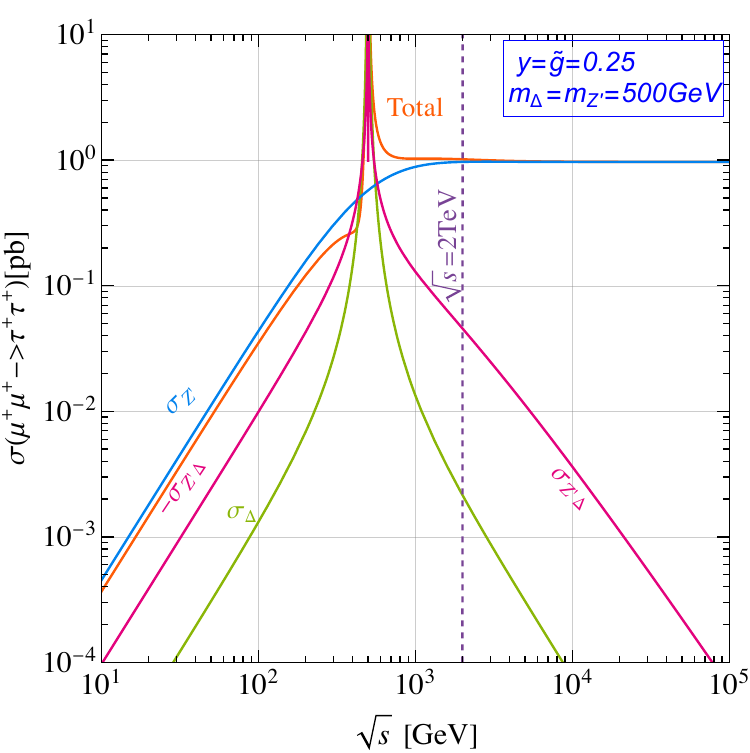}}
    \caption{The  cross section of $\mu^+\mu^+\to \ell^+\ell^+$ with $\sqrt{s}$. 
The contributions from individual components are shown in different colors. The center-of-mass energy of $\mu$TRISTAN, $\sqrt{s} = 2$ TeV, is indicated by the dashed line.
    (a) The left panel corresponds to the case of $\mu^+\mu^+\to \mu^+\mu^+$ with $y=0.25$ and $m_\Delta=500$ GeV.
    (b) The right panel means  the case of $\mu^+\mu^+\to \tau^+\tau^+$ with
  $y=\tilde g=0.25$, $m_\Delta=m_{Z'}=500$ GeV.  
   }
    \label{fig:tau}
\end{figure*}

The process $\mu^+\mu^+ \to \mu^+\mu^+$ conserves charged lepton flavor, implying the presence of SM background contributions. 
The total amplitude consists of three components: the SM contribution (mediated by $h$, $Z$, and $\gamma$ in the $t$- and $u$-channels), the new physics (NP) contribution (via $s$-channel exchange of $\Delta^{++}$), and the interference between them. 
Consequently, the total cross section comprises the pure SM, pure NP, and interference terms.
A direct calculation of the cross section reveals that $\sigma(t)$ exhibits an infrared divergence, particularly in the limit where the Mandelstam variable $t \to 0$ or $t \to -s$.
 This  is due to the fact that the interference  term includes one term proportional to $\ln(t/(t+s))$.
 In order to obtain the finite results, we can impose the cuts by considering  the bounds on rapidity $|\eta|<2.5$, 
 which is commonly used in the detector-level simulations.
 The rapidity has the relation with  variable t as
\begin{eqnarray}
    t=-\frac{s}{1+e^{-2\eta}}\to t\in \left[-\frac{s}{1+e^{-2\eta_{\text{max}}}},-\frac{s}{1+e^{2\eta_{\text{max}}}}\right]
\end{eqnarray}
Here we neglect the muon mass in this analysis. By setting the rapidity cut to $\eta_{\text{max}} = 2.5$, we compute the cross section for the benchmark values $y = 0.25$ and $m_\Delta = 500$ GeV, as shown in Fig.~\ref{fig:mumu}. 
The individual contributions are represented by different colors: the total cross section $\sigma_{\text{total}}$ (orange), the pure triplet contribution $\sigma_\Delta$ (green), the SM contribution $\sigma_{\text{SM}}$ (blue), and the SM–triplet interference term $\sigma_{\text{SM}-\Delta}$ (red).
We find that the SM contribution dominates across the full range of $\sqrt{s}$, except near the resonance at $\sqrt{s}\sim m_{\Delta}$, where the enhancement is due to the $s$-channel mediation of $\Delta^{++}$  resonance. 
The interference term is positive when $\sqrt{s} < m_\Delta$, and becomes negative for $\sqrt{s} > m_\Delta$. 
In particular, at $\sqrt{s} = 2$ TeV, the SM contribution significantly dominates, exceeding the sub-dominant interference term by approximately two orders of magnitude.

In order to extract signatures  in cases involving the SM background, we need to adopt the following criteria as
\begin{eqnarray}\label{crit}
CL= \frac{S}{\sqrt{S+B}}=\frac{\mathcal{L}|\sigma_{S}|}{\sqrt{\mathcal{L}(\sigma_{S}+\sigma_{B})}}=\sqrt{\mathcal{L}}\frac{|\sigma_{S}|}{\sqrt{(\sigma_{S}+\sigma_{B})}}\geq 3\;.
\end{eqnarray}
The relevant bounds are shown in blue in Fig.~\ref{fig:bound}. 
We find that there exist some parameter regions for $\Delta$ that simultaneously satisfy both phenomenological constraints and collider sensitivity, such as the benchmark point $y=0.25$, $m_\Delta=500$ GeV, and $v_\Delta=1$ GeV.

\subsubsection{$\mu^{+}\mu^{+} \rightarrow \tau^{+} \tau^{+}$}


For  $\mu^{+}\mu^{+} \rightarrow \tau^{+} \tau^{+} $,  it involves simultaneous charged lepton flavor violation and lepton number violation.
This feature implies the absence of SM background contributions, rendering it a promising collider signature at the two-body level within our model.
The flavor-changing $Z'$ model incorporating a triplet scalar can contribute to this process via two distinct mechanisms:
$\Delta^{++}$ mediated $s$-channel and 
  $Z'$ mediated $t$- and $u$-channels.
  The corresponding scattering amplitudes are
  \begin{eqnarray}
    M_{s}&=&\frac{1}{2}\frac{y^{2}}{s-m_{\Delta}^{2}+im_{\Delta}\Gamma_{\Delta}}\bar{\mu}(p_{1})\gamma^{\mu}P_{L}\tau(p_{4})\bar{\mu}(p_{2})\gamma_{\mu}P_{L}\tau(p_{3})\;,\\
    M_{t}&=&-\frac{\tilde{g}^{2}}{t-m_{Z'}^{2}+im_{Z'}\Gamma_{Z'}}\bar{\mu}(p_{1})\gamma^{\mu}\tau(p_{3})\bar{\mu}(p_{2})\gamma_{_\mu}\tau(p_{4})\;,\nonumber\\
    M_{u}&=&\frac{\tilde g^{2}}{u-m_{Z^{\prime}}^{2}+im_{Z^{\prime}}\Gamma_{Z^{\prime}}}\bar{\mu}(p_{1})\gamma^{\mu}\tau(p_{4})\bar{\mu}(p_{2})\gamma_{\mu}\tau(p_{3}) \nonumber \;.
  \end{eqnarray}
here the momenta of  initial state $\mu^+$ and final state $\tau^+$  are denoted as $p_{1,2}$ and $p_{3,4}$.
   The Mandelstam variables are denoted by $s$, $t$, and $u$ and Fierz transformations are applied.
Using FeynCalc~\cite{Shtabovenko:2023idz}, we calculate
 the cross section as the sum of the following three terms 
 \begin{eqnarray}\label{tautau}
  & &  \sigma(\mu^+\mu^+\to \tau^+\tau^+) =\sigma_{\Delta}+\sigma_{Z'}+\sigma_{Z'\Delta}\;,\\
   & & \sigma_{\Delta} =\frac{y^4}{64\pi }\frac{s }{\left(s-m_\Delta^2\right)^2+\Gamma_\Delta ^2 m_\Delta^2}\;,\nonumber\\
   & & \sigma_{Z'}  = \frac{ \tilde g^4
   }{4\pi s}\left[1+
  \frac{m_{Z'}^2\left(2
   m_{Z'}^2+3  s\right) }{s(2
   m_{Z'}^2+s)}\log \frac{m_{Z'}^2
   \left(\Gamma_{Z'}^2+m_{Z'}^2\right)}{
    \Gamma_{Z'}^2   m_{Z'}^2+\left(m_{Z'}^2+s\right)^2}
    +\frac{
   \left(m_{Z'}^4+m_{Z'}^2 \left(2 s-\Gamma_{Z'}^2\right)+2
   s^2\right)}{s\Gamma_{Z'} m_{Z'}}\tan ^{-1}\left(\frac{s\Gamma_{Z'}/m_{Z'}}{(s+m_{Z'}^2+\Gamma_{Z'}^2)}\right)\right],\nonumber\\
    & & \sigma_{Z'\Delta}  = \frac{y^2 \tilde g^2
   }{16\pi \left(s-m_\Delta^2\right)^2+\Gamma_\Delta ^2 m_\Delta^2}\left[
   2\Gamma_\Delta m_\Delta
   \tan ^{-1}\left(\frac{s\Gamma_{Z'}/m_{Z'}}{(s+m_{Z'}^2+\Gamma_{Z'}^2)}\right)
   -(s-m_\Delta^2)\log \frac{m_{Z'}^2
   \left(\Gamma_{Z'}^2+m_{Z'}^2\right)}{
    \Gamma_{Z'}^2   m_{Z'}^2+\left(m_{Z'}^2+s\right)^2}
   \right]\;.\nonumber
  \end{eqnarray}
  Here we ignore the charged lepton masses of both the initial and final states.
The symbols $\sigma_\Delta$ and $\sigma_{Z'}$ denote the $s$-channel $\Delta$ and the $t$- and $u$-channel $Z'$ contributions, respectively, while $\sigma_{Z'\Delta}$ represents the interference term between $\Delta$ and $Z'$.
Additionally, in certain scenarios, the interference term can be negative, particularly when $\sqrt{s} < m_\Delta$.
Analyzing the structure of the cross section, we find that in certain limits it approximates to
  \begin{eqnarray}\label{appro}
\sigma (\mu^+\mu^+\to \tau^+\tau^+) \propto
\begin{cases}
    \frac{y^4}{\Gamma_\Delta^2}, & \text{for } \sqrt{s} \simeq m_\Delta, \\\\
    \frac{\tilde g^4}{m_{Z'}^2}, & \text{for } \sqrt{s} \gg m_{Z'}, \\\\
    \tilde g^4\frac{s}{m_{Z'}^2}, & \text{for } \sqrt{s} \ll m_{Z'}, 
\end{cases}
  \end{eqnarray}

The analytic expressions of the cross section given in Eq.(\ref{tautau}) as a function of $\sqrt{s}$, including its different components, are illustrated in Fig.\ref{fig:tautau}.
Here, we set $y = \tilde{g} = 0.25$ and $m_\Delta = m_{Z'} = 500$ GeV.
We find that at large $\sqrt{s}$, the dominant contribution arises from $Z'$, which exceeds the $\Delta$ contribution by approximately two orders of magnitude and the interference terms by about three orders of magnitude.
The $\Delta$ contribution dominates only when $\sqrt{s} \sim m_\Delta$, as indicated in Eq.~(\ref{appro}).
Additionally, for $\sqrt{s} < m_\Delta$, negative interference terms arise, partially canceling contributions from the pure $Z'$ and $\Delta$ channels.
Fortunately, for the chosen benchmark points and at $\sqrt{s} = 2$ TeV, the negative interference scenario does not occur.
Requiring the event numbers  $N=\sigma\cdot \mathcal{L}\geq 3$, we can obtain the corresponding sensitivity  indicated by the red regions in Fig.~\ref{fig:all}.

\subsection{Four-body final states }

Considering the prompt decays of $Z'$ and $\Delta^{++}$, the produced $Z'$ and $\Delta^{++}$ will undergo subsequent decays rapidly.
Pair production occurs only at opposite-sign muon colliders $\mu^+\mu^-$, which has been studied in Ref.~\cite{Sun:2023ylp}.
Therefore, we primarily focus on single production at $\mu$TRISTAN followed by subsequent prompt decays, particularly targeting four visible charged-lepton final states exhibiting clear lepton flavor and generation number violation.
These  processes four-body final states include 
 $\mu^+ \mu^+ \to  \mu^+\tau^+ Z'(\to \tau^+\mu^-)$, $\mu^+ \mu^+ \to  \mu^+\mu^- \Delta^{++}(\to \tau^+\tau^+)$, 
 $\mu^+ \mu^+ \to  \mu^+\mu^-  \tau^+\tau^+$,  $e^- \mu^+ \to  e^- \tau^+ Z'(\to \tau^+\mu^-)$,
 $e^- \mu^+ \to  e^- \mu^- \Delta^{++}(\to \tau^+\tau^+)$
 and 
$e^- \mu^+ \to  e^- \mu^-  \tau^+\tau^+$.

 \begin{table*}[htbp!]
\centering
\caption{ The signal event numbers for different collider processes. Here we choose $y=0.25$ and $m_\Delta=500$GeV. The symbol with ``-" denotes kinematically forbidden processes.
}
\begin{tabular}{|c|c|c|c|c|c|c|c|c|c|c|c|c|c|}
\hline \multirow{3}{*}{Facility} & \multirow{3}{*}{$\sqrt{s}(\mathrm{GeV})$} & \multirow{3}{*}{$\mathcal{L}_{lum}(\mathrm{ab}^{-1})$}& \multirow{3}{*}{Process}
& \multicolumn{4}{|c|}{Event numbers}\tabularnewline\cline{5-8} 
&&&& \multicolumn{2}{|c|}{$\tilde g=0.05$} & \multicolumn{2}{|c|}{$\tilde g=0.3$} \tabularnewline\cline{5-8}  
& & & & $m_{Z'}=200$GeV & $m_{Z'}=500$GeV &  $m_{Z'}=200$GeV & $m_{Z'}=500$GeV \\\hline
$\mu^+\mu^+$ & 2000 & 0.012  &
\begin{tabular}{c}
$\tau^+\tau^+$ \\
$\mu^+\tau^+Z'(\to \tau^+\mu^-)$ \\
$\mu^+\mu^-\Delta^{++}(\to \tau^+\tau^+)$ \\
$\mu^+\mu^- \tau^+\tau^+$ \\
\end{tabular}&
\begin{tabular}{c}
3.70 \\
22.13\\
 111.58\\
0.15\\
\end{tabular}&
\begin{tabular}{c}
2.18 \\
 8.82\\
112.22 \\
0.12\\
\end{tabular}&
\begin{tabular}{c}
 8163.75\\
12420.8\\
111.88\\
270.72\\
\end{tabular}&
\begin{tabular}{c}
191.93 \\
3007.33 \\
112.03 \\
156.45 \\
\end{tabular}\\
\hline  $\mu^+e^-$ & 346 & 1&
\begin{tabular}{c}
$e^-\tau^+ Z'(\to \tau^+\mu^-)$ \\
$e^-\mu^- \Delta^{++}(\to \tau^+\tau^+)$ \\
$e^-\mu^- \tau^+  \tau^+$ \\
\end{tabular}&
\begin{tabular}{c}
 5.99\\
 -\\
 1.68\\
\end{tabular}&
\begin{tabular}{c}
 -\\
 -\\
 0\\
\end{tabular}&
\begin{tabular}{c}
7791.54 \\
-\\
5689.69 \\
\end{tabular}&
\begin{tabular}{c}
 -\\
-\\
2.08\\
\end{tabular}  \\
\hline
\end{tabular}
\label{number}
\end{table*}

Given the complexity of the four-body final state, we utilize MadGraph to perform a detailed collider analysis.
First, we implement the model interactions using FeynRules~\cite{Alloul:2013bka} to generate a Universal FeynRules Output (UFO) model~\cite{Degrande:2011ua} based on the model Lagrangian.
Then, we feed the model into MadGraph5-aMC@NLO~\cite{Alwall:2011uj} for all simulations, using PYTHIA 8~\cite{Sjostrand:2014zea} for parton showering and hadronization, and DELPHES~\cite{deFavereau:2013fsa} for fast detector simulation.
The  following basic cuts are also imposed:
(i) transverse momentum  $p_{T}>10$ GeV, 
(ii) absolute pseudo-rapidity $|\eta|<2.5$,
(iii) the separation  of the two leptons $\Delta R=\sqrt{(\Delta \eta)^2 +(\Delta \phi)^2}>0.4$.
The relevant signal event numbers for various collider processes are presented in Table~\ref{number}.

The collider analysis for four-body final states is organized as follows:
\begin{itemize}
 \item \textbf{$\mu^+ \mu^+ \to  \mu^+\tau^+ Z'(\to \tau^+\mu^-)$}:  It involves the on-shell single emission of a $Z'$ boson from one of the initial-state $\mu^+$, followed by its subsequent decay.
 The decays into $\mu^+\tau^-$ and $\mu^-\tau^+$ have equal branching ratios.
For simplicity, we focus primarily on the decay channel $Z' \to \tau^+ \mu^-$, due to its charged lepton flavor violation signature.
\item \textbf{$\mu^+ \mu^+ \to  \mu^+\mu^- \Delta^{++}(\to \tau^+\tau^+)$}: 
One of the initial-state $\mu^+$ emits an on-shell $\Delta^{++}$, which decays promptly.
Similarly, the decays into muon and tau flavors share identical branching ratios.
However, only the decays into tau flavors violate charged lepton flavor conservation.
\item \textbf{$\mu^+ \mu^+ \to  \mu^+\mu^-\tau^+  \tau^+$}: 
This process clearly violates charged lepton flavor conservation.
Since there are no intermediate on-shell states, the process receives contributions from both $\Delta$ and $Z'$, whether on-shell or off-shell.
\item \textbf{$e^- \mu^+ \to  e^- \tau^+ Z'(\to \tau^+\mu^-)$}: 
It involves the on-shell single emission of a $Z'$ boson from $\mu^+$, followed by its decay into $\mu^- \tau^+$, exhibiting charged lepton flavor violation.
Note that on-shell production requires $m_{Z'} < \sqrt{s}$, meaning the process only probes mass regions below the COM $\sqrt{s}$.
\item \textbf{ $e^- \mu^+ \to  e^- \mu^- \Delta^{++}(\to \tau^+\tau^+)$}:
On-shell production of $\Delta$ is forbidden since its lower mass bound, $m_\Delta > 420$ GeV, exceeds the center-of-mass energy $\sqrt{s} = 346$ GeV.
\item \textbf{$e^- \mu^+ \to  e^- \mu^-  \tau^+\tau^+$}: 
This process does not involve any on-shell intermediate states, meaning it is influenced by off-shell $\Delta^{++}$ and either on-shell or off-shell $Z'$. The combined effects of both contribute jointly to the process.	
\end{itemize}

\begin{figure}
    \centering
\includegraphics[width=1.\linewidth]{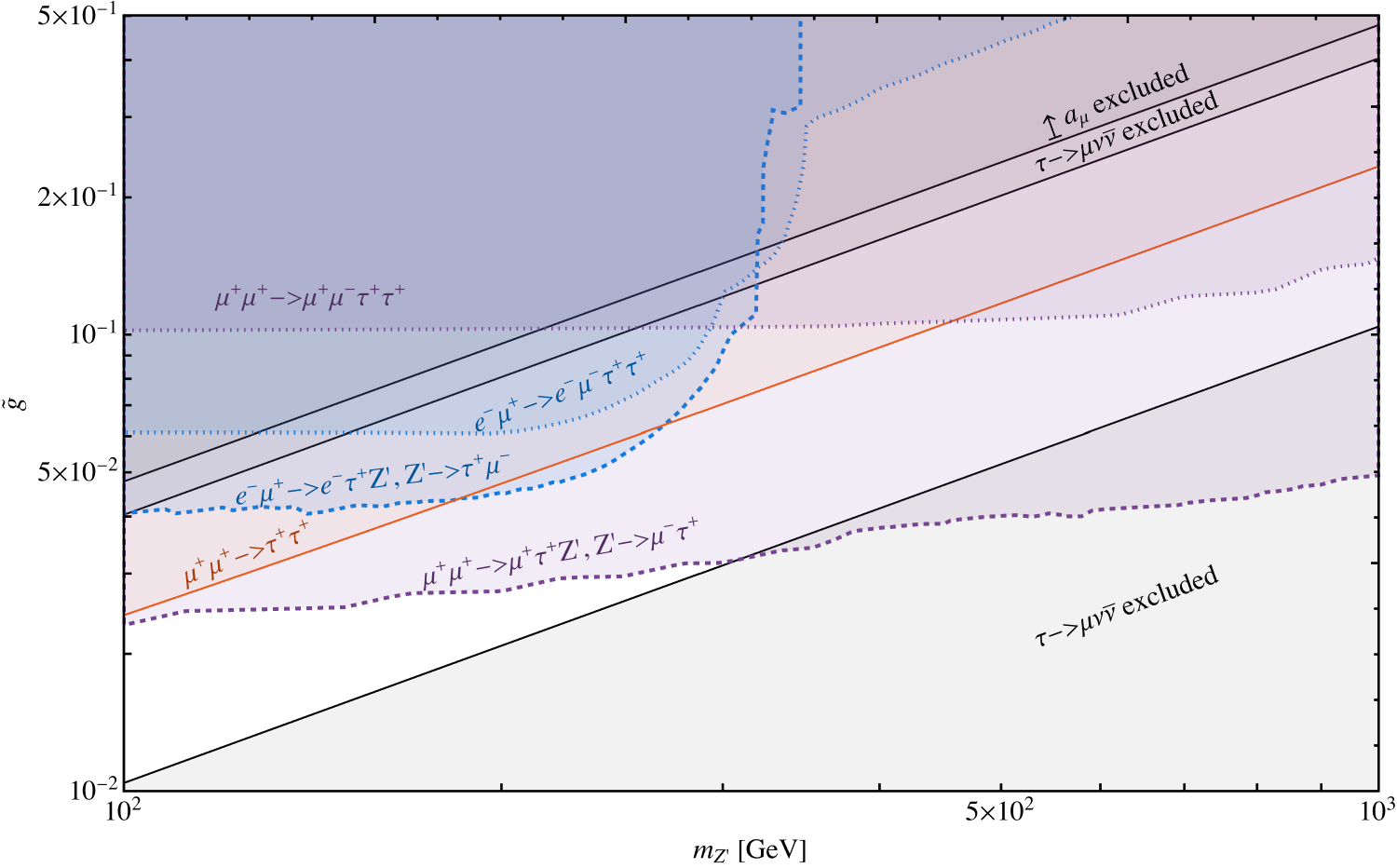}
    \caption{
    Current experimental bounds and future $\mu$TRISTAN sensitivities  in the $m_{Z'}$–$\tilde{g}$ plane. The benchmark values are fixed at $y=0.25$ and $m_\Delta = 500$ GeV.
    The  gray areas indicate exclusion limits derived from $(g-2)_\mu$\cite{Muong-2:2025xyk,Aliberti:2025beg} and $\tau \to \mu \nu \bar{\nu}$\cite{Cheng:2021okr,Huang:2024iip}.
 Different collider processes are depicted in distinct colors, with on-shell contributions (dashed) and processes without intermediate states (dotted).
    }
    \label{fig:all}
\end{figure}

The relevant findings are illustrated in Fig.~\ref{fig:all}, 
showing both current bounds (light gray regions) and projected sensitivities of $\mu$TRISTAN colliders (colored regions).
The gray regions summarize current experimental constraints, as mentioned earlier, including muon $(g-2)_\mu$ and $\tau\to \mu \nu\bar\nu$.
We find that the decay $\tau \to \mu \nu \bar{\nu}$ provides significantly more stringent bounds than the muon $(g-2)_\mu$.
Although a large portion of the parameter space has already been excluded,  some viable regions remain to be explored at colliders.

The colored regions represent the projected sensitivities from various collider processes at $\mu$TRISTAN. These are derived by requiring the number of events $N=\sigma\cdot \mathcal{L}>3$.
Our analysis demonstrates that these processes can probe new parameter regions that remain unconstrained by current experimental bounds.
For the two-body process $\mu^+\mu^+ \to \tau^+\tau^+$, the coupling $\tilde{g} \sim 0.025$ can be probed, which improves sensitivity by approximately a factor of two compared to the bounds from $\tau \to \mu \nu \bar{\nu}$.
For on-shell $Z'$ production, indicated by the dashed line, the process $e^- \mu^+ \to e^- \tau^+ Z' (\to \tau^+ \mu^-)$ can achieve sensitivity around $\tilde{g} \sim 0.04$ for $m_{Z'} < 346$ GeV, while $\mu^+ \mu^+ \to \mu^+ \tau^+ Z' (\to \tau^+ \mu^-)$ can reach down to $\tilde{g} \sim 0.024$.
Moreover, $\mu^+ \mu^+ \to \mu^+ \tau^+ Z' (\to \tau^+ \mu^-)$ provides significantly better sensitivity across the entire mass range, fully covering the regions probed by both $e^- \mu^+ \to e^- \tau^+ Z' (\to \tau^+ \mu^-)$ and $\mu^+ \mu^+ \to \tau^+ \tau^+$.
In scenarios without intermediate states,  $\mu^+\mu^+ \to \mu^+\mu^-\tau^+\tau^+$, the coupling can reach $\tilde g \sim 0.1$ across the entire mass region, which is fully covered by the previous scenario involving an on-shell $Z'$.
This suggests the presence of negative interference between the $Z'$ and $\Delta$ contributions, which significantly reduces the cross section.

\section{Conclusion}
\label{sec4}

In this work, we investigate the detection prospects of the $U(1)_{L_\mu-L_\tau}$ model with triplet scalars $\Delta$ at the $\mu$TRISTAN collider.
The model framework features maximal flavor-changing $U(1)_{L_\mu-L_\tau}$ $Z'$ interactions alongside flavor-conserving $\Delta$ interactions.
Both contribute to the muon  $(g-2)_\mu$ and the decay $\tau \to \mu \nu \bar{\nu}$ with opposite effects, which can explain the recent absence of $(g-2)_\mu$ anomaly while satisfying the stringent constraints from tau decay.
These counteracting effects render the model phenomenologically interesting and warrant further collider search analyses.
Given that the same-sign muon collider offers advantages over the opposite-sign muon collider, we focus on the sensitivity of $\mu$TRISTAN through various charged lepton flavor violation processes.
We find that for $m_{Z'}$ in the hundreds of GeV range, the $\mu^+\mu^+$ and $\mu^+ e^-$ collider modes at $\mu$TRISTAN can probe parameter regions inaccessible to current experiments and offer greater projected sensitivity than opposite-sign muon colliders, especially for processes like  $\mu^+ \mu^+ \to  \mu^+\tau^+ Z'(\to \tau^+\mu^-)$. 
This suggests that $\mu$TRISTAN can serve as a complementary probe of the $U(1)_{L_\mu-L_\tau}$ model, providing strong motivation for the next generation of high-energy lepton colliders.

\section*{Acknowledgments}
 The work of Fei Huang is supported by Natural Science Foundation of Shandong Province under Grant No.ZR2024QA138, and
 Jin Sun is supported by IBS under the project code, IBS-R018-D1..

\bibliographystyle{JHEP}
\bibliography{ref}

\end{document}